# Integrating economic and psychological insights in binary choice models with social interactions

Katarzyna Ostasiewicz[1], Michal H. Tyc[1], Piotr Goliczewski[1], Piotr Magnuszewski[1], Andrzej Radosz[1], Jan Sendzimir[2]

[1] – Institute of Physics, Wroclaw University of Technology, Poland
[2] – International Institute for Applied Systems Analysis, Laxenburg, Austria

Abstract

We investigate a class of binary choice models with social interactions. We propose a unifying perspective that integrates economic models using a utility function and psychological models using an impact function. A general approach for analyzing the equilibrium structure of these models within mean-field approximation is developed. It is shown that within a mean-field approach both the utility function and the impact function models are equivalent to threshold models. The interplay between heterogeneity and randomness in model formulation is discussed. A general framework is applied in a number of examples leading to some well-known models but also showing the possibility of more complex dynamics related to multiple equilibria. Our synthesis can provide a basis for many practical applications extending the scope of binary choice models.

1. Introduction.

People are social animals. Decisions in many situations are strongly influenced by decisions other people make. In economic theory, people making choices are conceptualized as agents that, endowed with preferences and expectations about others' choices, decide in the face of constraints. Preferences are usually formalized using utility functions. Interactions between agents can be indirect (choices of some agents can induce constraints on others) or direct (choices can patently influence other's choices or expectations). Although it was argued that economic actions are embedded in structure of social relations [Granovetter 1985], direct social interactions were for a long time usually not incorporated into economic models. Part of the reason for this was analytical intractability – interaction term introduces complexity which is not easy to analyze mathematically. However, that very complexity can be responsible for the novel, emergent phenomena whose prospect may provide explanations of collective social phenomena [Durlauf 2005, Arthur 2006, Rosser 1999]. Recently, models with social interactions are gaining increasing attention.
There have been some isolated attempts (see e.g. [Veblen 1934, Simon 1954]) to account in economic theory for social interactions not mediated by market. However, more systematic exploration of these effects started with Schelling's model of racial segregation [Schelling 1971, Schelling 1978]. Agents in Schelling's model make decisions to stay or to leave a neighborhood based on some threshold levels of neighbors with the same race. The idea of threshold based decisions was also investigated by Granovetter in his model of collective behavior [Granovetter 1979]. These models spurred extensive research and modeling efforts on so called "threshold" or "critical mass" models (see e.g. [Marwell and Oliver 1993, Oliver and Marwell 2001, Abrahamson and Rosenkopf 1993, Akerloff 1980, Lindbeck 1997]) .
One increasingly popular approach to studying social interactions is based on an analogy with interacting particles using methods of statistical mechanics. In economics it was initiated by Folmer [Folmer 1974], but it proliferated in 1990s with the models proposed by Durlauf, Brock and Blume [Durlauf 1991, Durlauf 1993, Durlauf 1999, Brock, Durlauf, 2001a, Brock

Durlauf 2001b, Blume 1993, Blume 1995, Blume Durlauf 2003, Phan et al. 2004] belonging to the class of Random Utility Models (RUM) [McFadden 1974]. In these models interactions between agents are incorporated by adding a social term to individual utility functions. Another approach based on the voter model was proposed by Glaeser and coauthors [Glaeser et al 1996, Glaeser and Scheinkmann 2002]. These efforts are mostly based on the assumption that agents' decisions are based on some kind of an average choice of other agents in her/his surroundings (global or local). Such an assumption makes it natural to use some kind of mean-field theory [Brock 1993]. Richer, more heterogeneous 'social geometry' is much harder to tackle analytically [Ioannides 2006, Horst and Scheinkman 2004] and such models have been often treated through simulations (see e.g. [Phan 2003]).

Economists' efforts to apply models and methods from physics have been reflected by physicists seeking economic applications of their models. Many attempts to apply statistical physics tools to analyze social systems focused on opinion dynamics as influenced by social interactions [Weidlich and Haag 1983, Galam et al. 1982, Galam 1997, Galam and Zucker 2000, Sznajd-Weron and Sznajd 2000, Deffuant et al. 2000, Holyst et al. 2000]. These models can be applied to other binary choice situations studied in economics – one of the differences with economic models is that agents usually do not hold expectations but make decisions based on actual states of other agents.

Social interactions have been studied intensively in social psychology. However, the lack of formal rigor and common conceptualization has generated many overlapping ideas whose fuzziness complicates the building of formal models. One notable line of models was developed using social impact theory by Nowak, Szamrej and Latane [Nowak et al. 1990]. Social impact theory [Latane 1981], formulated initially in the static setting of a group impact on an individual, is backed by considerable empirical evidence [Latane and Nida 1981, Freeman et al. 1975, Latane and Williams 1979, Jackson and Latane 1981, Latane and Wolf 1981]. The theory states that social impact exerted by the group on an individual is proportional to the "strength" of interaction, social distance and number of group members. Additionally two types of interactions are distinguished: supporting an agent in her/his attitude (it can also be an opinion or decision) and persuading him/her to change an attitude. Application of the theory to the group setting using computer models [Nowak and Lewenstein 1996, Latane and Nowak 1994] led to interesting results. In particular, the model was able to reproduce the survival of minority clusters in equilibrium. Another interesting feature of this model is inclusion of 'self-support' – a tendency of an individual to sustain her/his opinion. This means that an agent's choice depends on a choice she/he formerly made, which produces a certain inertia in the agent's behavior resulting from the psychological tendency toward consistent [Festinger 1957, Aronson 1992, Cialdini 2001] or habitual behavior (it can also be interpreted as an agent's susceptibility to outside influence). Results generated by these models inspired a new generation of models and applications [Holyst et al. 2000, Holyst et al. 2001, Nowak et al. 2000] within emerging dynamic social psychology [Kenrick et al. 2003, Vallacher et al. 1994, Vallacher et al. 2002].

Models of social interaction have often been investigated for the existence and uniqueness of systems equilibria, a central idea of economic models analysis. The presence of social pressure from a peer group can change structural properties of discrete choice models, producing multiple equilibria. One of them is a global equilibrium; others correspond to local ones. Existence of multiple equilibria brings important consequences for system dynamics – such models can undergo sudden and fast changes in their macroscopic properties within a small range of parameters [Holyst et al. 2000, Brock 2004, Levy 2005, Gligor and Ignat 2001]. In a manner analogous to phase transitions phenomena in physics, some authors call them "social phase transitions," although one should be careful with extending this analogy too far[1]. Closely related to this is the "hysteresis effect" where the sequence of states the system goes through with a slow parameter change is different when changing this parameter in the opposite direction, forming a closed loop. Such behavior can be a reason for often slow responses of groups to threats or opportunities [Scheffer et al. 2003]. Systems with such features can be used to represent 'social traps' [Platt 1973] where a system stays in a globally undesired configuration although all agents follow their preferences.

Social interaction models have been examined in numerous applications including: opinion dynamics, technology choice, rumor spread, strikes, voting, spatial agglomeration, social pathologies, information cascades etc. [Brock and Durlauf 2001b, Durlauf 1997, Kirman 1997, Zanella 2004, Goldstone and Janssen 2005]. They have been used extensively to model the role of 'neighborhood effects' in determining socioeconomic outcomes [Sampson et al. 2002, Durlauf 2003]. Although a plethora of possible applications have been claimed, links with empirical evidence are not always strong. Actually serious methodological problems arise in attempts to identify causal connections between social interactions and social phenomena under study. These identification and selection problems result from difficulties in distinguishing between endogenous effects (direct social interaction), contextual effects (agents are influenced by characteristics of other agents) and correlated effects (agents share common unobserved factors) [Manski 1993, Brock and Durlauf 2001b, Glaeser et al 1996, Ioannides and Zabel 2003, Ioannides 2006, Heckman 2001].

Modeling efforts in economics, sociology and social psychology have been so far developed to a large extent in isolation. One recent effort to unify these different approaches came from Levy [Levy 2005], who proposed a framework to integrate utility models with threshold models. In this article we focus on a perspective that integrates binary choice models with direct (endogenous) social interactions. The variable representing each agent's choice can also be interpreted as an attitude or an opinion, an interpretation more common in social psychology models. We assume that agents have direct information about other agents' choices, i.e. myopic expectations. Such agents operate on what they directly know, not what they expect. This means also that we do not assume that agents posses any structural knowledge about the system they operate in (e.g. equilibria of the system or distribution of other agent's choices).

This paper is organized as follows. In Section 2 we propose a general framework for individual-based binary choice, social interaction models. We show that under the assumption that an agent's choice depends on her/his previous choice, the economic models using the utility function and social psychology models using the impact function are mathematically equivalent. This equivalence allows us to use insights from both economics and social psychology to develop models applicable to a wider range of situations. Economic models with social interactions already expanded the limited *homo economicus* assumptions of neoclassical theory. Utility of agents is not only based on economic benefits and costs but also includes gains and losses related to conforming with others' choices. Psychological perspectives add inertia to agents' decisions, a certain tendency to follow the last choice. Furthermore Nowak and Latane [Nowak et al. 1990] proposed to disaggregate the social interaction term into the two parts: supporting (sustaining the last choice) and persuading (pushing for change from the last choice). In general these social interaction terms can be nonlinear functions of the number of supporting or persuading agents. Nonlinearity of these functions allows us to model more realistic assumptions about agents' decision rules and can result in a richer equilibrium structure.

---

[1] In fact, "social phase transitions" may be associated with the category of the second order phase transitions in statistical physics [Landau, Lifshitz, 1980]. There are two key features related to these phenomena. The first one is degeneration of the ground state. The second one is a universal behavior of different systems leading to a division into classes of universalities characterized by so-called critical exponents. In social systems with a binary choice, the former condition is fulfilled itself: 'yes' or 'no' decisions of all agents correspond, at least in principle, to the "two degenerated ground states of the system". The question of universal behavior turns out to be much more subtle: statistical physics phase transitions are equilibrium phenomena, with a well defined meaning of a temperature as an obvious background related to one of the fundamental laws, namely maximal entropy [Landau, Lifshitz, 1980]. In social science an attempt to define "the temperature" and/or "an equilibrium" and, in consequence, the meaning of "social temperature" or "social equilibrium" is not as well established a process as it is in statistical physics due to the lack of such a firm ground as the above mentioned law of nature.

In Section 3 we introduce the so-called mean field approximation, where any individual choice depends only on the average choice of all other agents. Unlike other authors [Durlauf 1999, Phan et al. 2004] we do not explore the link with statistical physics directly, which allows us to obtain results for a larger class of models. Although mean-field approximation does not allow us to include the *geometry* of social relationships (social network), it becomes a significant reference model. It allows us to distinguish specific aspects of social space geometry which can generate different results. Within a mean-field approach both utility function models and social impact models can be formulated as 'threshold' models. Such conceptualization, first introduced by Granovetter, was recently developed by Levy [Levy 2005] who showed its equivalence with mean field utility function models. However both Levy and Granovetter considered transitions of individuals' choices only in one direction (adoption), such that reversing a decision is impossible. It is a very useful approach for problems like innovation diffusion. We consider here a more general framework where transitions can occur in both directions. In such cases the simple graphical analysis used by Granovetter and Levy to obtain systems equilibria does not hold. We propose here a general procedure for analyzing equilibria in such models. This procedure will be applied in Section 5 to specific cases.

In Section 4 we propose a generalized model for the binary choice problem. In this model one takes the utility function in a certain form: it is assumed to be an additive function of individual preferences, inter-personal influences, randomness and 'self-supportiveness' (individuals' inertia). Individual preferences can result from personal characteristics or can be a result of external influence. In the physics of phase transitions this component is called the 'external field'. With these assumptions we obtain a more specific condition for systems equilibria. Also we discuss relation between some forms of randomness in individuals' choices and heterogeneity of agents' characteristics. As we relax some of the assumptions of the existing social interaction models (e.g. the form of the individual utility random term, heterogeneity conditions on some parameters) our results contribute to the synthesis of the field and expand possible applications.

In Section 5 there are given some examples of how, using specific assumptions, many binary choice models known from literature can be obtained from the generalized model proposed in Section 4. We also show how different assumptions about the form of social interactions can produce different equilibrium structures. Specifically, variety different equilibria (we have presented an example with five different possibilities) can be obtained in one of the versions of Nowak-Latane model.

We conclude in Section 6 with discussion and summary.

2. 'Individual' approach.

We propose here a general framework to construct models describing a group of individuals, between each pair of which may exist an interaction of various strength. At each time step every individual has to make a binary decision, which we denote here by (+1) ('yes', joining a certain action, etc.) or (-1) ('no', refusal of action, etc.). Individuals are assumed to choose an action with higher utility, which depends not only on individual preferences but also on others' choices. They make their choices dependent upon gains and losses which they evaluate for themselves, including social gains or losses from being a part of a group or an outsider. Moreover, random factors may also play a role in the decision making process. The current choice of an individual can also depend on the previous choice of that same individual (*inertia or habitual behavior of agents*).

In order to describe such a situation a *utility function* [Brock, Durlauf 2001b, Brock 2004, Durlauf 1999] can be defined as depending on a variety of variables and parameters:

$$U_i \equiv U_i\left(\sigma_i', \sigma_i, \{\sigma_{j \neq i}\}, \{\{q_{ij}\}\}, \{r_i\}, \varepsilon_i(\sigma_i')\right), \qquad (1)$$

where $\sigma_i'$ denotes the present choice of an individual $i$, $\sigma_i$ - his or her previous choice, $\{\sigma_{j \neq i}\}$ - choices of all the others individuals, $\{\{q_{ij}\}\}$ - set of interaction parameters (double brackets mean a set of sets: there may be a few parameters $\{q_{ij}\}$ referring to the given pair $ij$), $\{r_i\}$ - set of individual parameters describing his or her characteristics, and $\varepsilon_i(\sigma_i')$ - random term. $\varepsilon_i(+1)$ and $\varepsilon_i(-1)$ are random variables with some distributions. This term can be interpreted as a random component influencing all individual decisions. We shall consider $\varepsilon_i(\sigma_i')$ as an additive term, and only the distribution of difference $\varepsilon_i(-1) - \varepsilon_i(+1)$ will play the role.

An individual is assumed to make a decision comparing the values of $U_i\left(+1, \sigma_i, \{\sigma_{j \neq i}\}, \{\{q_{ij}\}\}, \{r_i\}, \varepsilon_i\right)$ and $U_i\left(-1, \sigma_i, \{\sigma_{j \neq i}\}, \{\{q_{ij}\}\}, \{r_i\}, \varepsilon_i\right)$, and choosing the state corresponding to the larger of these two values.

The other approach to that decision making problem is based on the so-called *impact function*. Within this approach the current decision is dependent on the previous choice of the individual in a distinguished manner. Current choice is defined via former one and impact function as follows [Nowak *et al* 1990, Lewenstein *et al* 1992, Holyst *et al* 2000]:

$$\sigma_i' = -\text{sgn}[\sigma_i I_i], \qquad (2)$$

where $\sigma_i'$ and $\sigma_i$ denote, as above, the present and the previous choice of an individual, respectively.

Impact function depends on the previous choice of the individual him or herself, as well, as on the choices of other individuals, some set of other parameters and random term. As one can see in Eq. (2), an individual's decision on changing or not changing former choice depends on the sign of impact function: positive (negative) value of impact function corresponds to a decision "changing" ("not changing").

It may be shown, that these two approaches are equivalent, i.e. having some utility function, one can calculate an impact function and visa-versa.

In fact, starting with the impact function one may notice, that:

a)  in the case $\sigma_i I_i < 0$, (in which according to (2) the new state, $\sigma_i'$ will take value (+1)), the value of expression,

$$-\sigma_i \sigma_i' I_i$$

is positive for $\sigma_i' = 1$ and negative for $\sigma_i' = -1$;

b) in the case $\sigma_i I_i > 0$, the value of expression,

$-\sigma_i \sigma_i' I_i$

is positive for $\sigma_i' = -1$ and negative for $\sigma_i' = 1$.

Thus, the new state, $\sigma_i'$, equals +1 or -1, when $-\sigma_i \sigma_i' I_i$ is positive or negative, respectively. Therefore, a utility function may be defined as follows:

$$U(\sigma_i') \equiv -\sigma_i \sigma_i' I_i. \qquad (3)$$

On the other hand, by definition of the utility function:

$$\sigma_i' = \text{sgn}[U_i(\sigma_i' = +1) - U_i(\sigma_i' = -1)] \Rightarrow \sigma_i' = -\text{sgn}\left[\sigma_i \frac{-(U_i(\sigma_i' = +1) - U_i(\sigma_i' = -1))}{\sigma_i}\right].$$

Thus, for arbitrary utility function, since $\sigma_i^2 = 1$, we can define a corresponding impact function

$$I_i = -\sigma_i (U_i(\sigma_i' = +1) - U_i(\sigma_i' = -1)), \qquad (4)$$

Therefore, these two types of approaches in modeling discrete choice problems are explicitly equivalent.

3. Mean-field approach.

In order to investigate the stationary properties of the system, the so-called mean-field approach is usually considered. In this approximation one can assume that all individuals interact with all the others with the same strength. That is, each individual depends on the average choice of all individuals and not on each of the others separately. Therefore, we will replace sets of parameters $\{\{q_{ij \neq i}\}\}$ by uniform (the same for all agents) coefficients $\{q\} = \left\{\frac{1}{N^2} \sum_{ij \neq i} q_{ij}\right\}$. The remaining parameter (or set of parameters) $\{q_{ii}\}$ is connected with the so-called *self supportiveness*, which measures the strength of dependence of the current choice on the previous choice of an individual.

Within the mean-field approach we will also assume that the number of individuals, $N$, is large enough to make

$$\frac{1}{N} \sum_{j \neq i} \sigma_j \approx \frac{1}{N} \sum_j \sigma_j.$$

The right-hand side of the above equation is a definition of a 'mean choice', denoted by $m$:

$$m \equiv \frac{1}{N} \sum_j \sigma_j, \quad m \in \langle -1, 1 \rangle.$$

In the framework of such a mean-field approximation both the *utility function model* and the *impact function model* can be reformulated as a *threshold model* [Granovetter 1979, Levy 2005].

*Threshold model.*

Having the utility function, $U_i \equiv U_i(\sigma_i', \sigma_i, m, \{q\}, \{q_{ii}\}, \{r_i\}, \varepsilon_i)$, one may calculate the condition for $m$, under which the transition from the state (-1) to the state (+1) or from the state (+1) to the state (-1) (conditional probabilities) will occur. We will denote $U_i(\sigma_i', +1, m, \{q\}, \{q_{ii}\}, \{r_i\}, \varepsilon_i)$ by $U_i^+(\sigma_i', m, \{q\}, \{q_{ii}\}, \{r_i\}, \varepsilon_i)$ and $U_i(\sigma_i', -1, m, \{q\}, \{q_{ii}\}, \{r_i\}, \varepsilon_i)$ by $U_i^-(\sigma_i', m, \{q\}, \{q_{ii}\}, \{r_i\}, \varepsilon_i)$. Thus the utility function may be written as:

$$U_i(\sigma_i', \sigma_i, m, \{q\}, \{q_{ii}\}, \{r_i\}, \varepsilon_i) = \frac{1}{2}\left[(1+\sigma_i)U_i^+(\sigma_i', m, \{q\}, \{q_{ii}\}, \{r_i\}, \varepsilon_i) + (1-\sigma_i)U_i^-(\sigma_i', m, \{q\}, \{q_{ii}\}, \{r_i\}, \varepsilon_i)\right] \quad (5)$$

and conditional probability is:

$$P(\sigma_i' = +1 | \sigma_i = -1) = P\left(\left[U_i^-(+1, m, \{q\}, \{q_{ii}\}, \{r_i\}, \varepsilon_i(+1)) - U_i^-(-1, m, \{q\}, \{q_{ii}\}, \{r_i\}, \varepsilon_i(-1))\right] > 0\right). \quad (6a)$$

Provided that for any realization of random variables $\varepsilon_i(\pm 1)$ holds:

$$\frac{d}{dm}\left[U_i^-(+1, m, \{q\}, \{q_{ii}\}, \{r_i\}, \varepsilon_i(+1)) - U_i^-(-1, m, \{q\}, \{q_{ii}\}, \{r_i\}, \varepsilon_i(-1))\right] > 0$$

and there exists such value of $m$, $m_i^{Th-}$, that satisfies:

$$U_i^-(+1, m_i^{Th-}, \{q\}, \{q_{ii}\}, \{r_i\}, \varepsilon_i(+1)) - U_i^-(-1, m_i^{Th-}, \{q\}, \{q_{ii}\}, \{r_i\}, \varepsilon_i(-1)) = 0,$$

one may replace Eq. (6a) with the new one, for the new random variable $m_i^{Th-}$:

$$P(\sigma_i' = +1 | \sigma_i = -1) = P(m_i^{Th-} < m) \equiv F_i^{Th-}(m). \quad (6a')$$

Analogously, provided that for any realization of random variables $\varepsilon_i(\pm 1)$ holds:

$$\frac{d}{dm}\left[U_i^+(-1, m, \{q\}, \{q_{ii}\}, \{r_i\}, \varepsilon_i(-1)) - U_i^+(+1, m, \{q\}, \{q_{ii}\}, \{r_i\}, \varepsilon_i(+1))\right] < 0$$

and there exists such value of $m$, $m_i^{Th+}$, that satisfies:

$$U_i^+(-1, m_i^{Th+}, \{q\}, \{q_{ii}\}, \{r_i\}, \varepsilon_i(-1)) - U_i^+(+1, m_i^{Th+}, \{q\}, \{q_{ii}\}, \{r_i\}, \varepsilon_i(+1)) = 0$$

The conditional probability of 'switching' opinion in the opposite direction may be written as:

$$P(\sigma_i' = -1 | \sigma_i = +1) = P\left(\left[U_i^+(-1, m, \{q\}, \{q_{ii}\}, \{r_i\}, \varepsilon_i(-1)) - U_i^+(+1, m, \{q\}, \{q_{ii}\}, \{r_i\}, \varepsilon_i(+1))\right] > 0\right) =$$
$$= P(m_i^{Th+} > m) \equiv 1 - F_i^{Th+}(m) \quad (6b)$$

$F_i^{Th-}(m)$ and $F_i^{Th+}(m)$ are cumulative distribution functions for the values of $m$ needed to precipitate a state change for individual $i$ ($m^{Th}$). These functions result from probability distributions for random terms $\varepsilon_i$ (in the special case, $\varepsilon_i \equiv 0$, $F_i^{Th-}(m)$ and $F_i^{Th+}(m)$ are step functions).

In order to find system equilibria within the mean-field approach, one needs 'collective' distribution functions $F^{Th-}(m)$ and $F^{Th+}(m)$. In the simplest case one can assume that the random terms $\varepsilon_i$ for all individuals have these same distributions $F_i^{Th-}(m)$ and $F_i^{Th+}(m)$, they are simply: $F^{Th-}(m) = F_i^{Th-}(m)$ and $F^{Th+}(m) = F_i^{Th+}(m)$. In the general case they are given by:

$$F^{Th-}(m) = \frac{1}{N}\sum_i F_i^{Th-}(m) \quad \text{and} \quad F^{Th+}(m) = \frac{1}{N}\sum_i F_i^{Th+}(m).$$

It is a generalized threshold model, containing two sets of thresholds for two directions of transitions.

In some cases (see below, Section 5) reformulating the model by using thresholds, allows us for quick graphical analysis of stationary properties of the system.

*Equilibria within mean-field approximation.*

We will define an equilibrium state as such a state, that:

$m' = m$.

This definition corresponds to a thermodynamic equilibrium in statistical physics: there are very many microscopic states (an exponentially large number of states, $2^N$, in the most simple case of an Ising-like model, where N is the number of "individuals"-spins) corresponding to such a "macroscopic state". In this case we also do not require a 'microscopic' equilibrium, i.e. individuals may change their states from (+1) to (-1) and visa versa. Only the mean value of their choices has to be constant.

The probability of current choice equal (+1) is:

$$P(\sigma_i' = 1) = P(\sigma_i' = 1 | \sigma_i = 1)P(\sigma_i = 1) + P(\sigma_i' = 1 | \sigma_i = -1)P(\sigma_i = -1), \qquad (7)$$

where $i$ corresponds to a randomly chosen individual.
As:

$P(\sigma_i = +1) - P(\sigma_i = -1) = m$
$P(\sigma_i = +1) + P(\sigma_i = -1) = 1$

one has:

$$P(\sigma_i = +1) = \frac{1+m}{2}, P(\sigma_i = -1) = \frac{1-m}{2}. \qquad (8)$$

Substituting (8) and (6) into (7) one obtains:

$$\frac{1+m'}{2} = F^{Th+}(m)\frac{1+m}{2} + F^{Th-}(m)\frac{1-m}{2}.$$

Using the definition of the equilibrium, $m' = m$, one finds:

$$m = \frac{F^{Th+}(m) + F^{Th-}(m) - 1}{1 + F^{Th-}(m) - F^{Th+}(m)}, \tag{9}$$

or, defining $m_+ = P(\sigma_i = +1)$,

$$m_+ = \frac{F^{Th-}(m)}{1 + F^{Th-}(m) - F^{Th+}(m)}. \tag{9a}$$

One can see that the right-hand side of Eq. (9a) has all the properties of a cumulative distribution function. Indeed, $\lim_{m \to -\infty} \frac{F^{Th-}(m)}{1 + F^{Th-}(m) - F^{Th+}(m)} = 0$, $\lim_{m \to +\infty} \frac{F^{Th-}(m)}{1 + F^{Th-}(m) - F^{Th+}(m)} = 1$ and $\frac{d}{dm}\left[\frac{F^{Th-}(m)}{1 + F^{Th-}(m) - F^{Th+}(m)}\right] \geq 0$ as the following expression is smaller or equal to zero:

$[F^{Th+}(m) - 1]\frac{d}{dm}F^{Th-}(m) - F^{Th-}(m)\frac{d}{dm}F^{Th+}(m)$. That is, one may regard the right-hand side of Eq. (9a) as a cumulative distribution function of some 'effective thresholds', $F^{Th\,eff}(m)$ and:

$$m_+ = F^{Th\,eff}(m). \tag{9b}$$

For some special cases a simple analysis of the properties of the function $F^{Th\,eff}(m)$ (or, equivalently, the right-hand side of Eq. (9)) allows us to predict qualitatively the existence and number of possible equilibria (see Sec. V). Let us mention that such a condition in statistical physics, of the existence of multiple equilibria in some temperature range, is an indicator of a possible phase transition. The system itself chooses, more or less spontaneously, one of the equivalent equilibria – one can refer to that as a spontaneously broken symmetry.

4. Generalized model.

Here, we propose a possible, generalized form of a utility function, which covers, as special cases, many utility/impact function models from literature, but is not limited to them. The proposed form imposes fewer restrictions than existing models and provides a richer variety of possible specifications.
We propose for the functions $U_i^+$ and $U_i^-$ in the following form (see also [Brock, Durlauf 2001b, Brock 2004, Durlauf 1999, Nowak et al 1990, Holyst et al 2000]):

$$U_i^\pm = \sigma_i' h_i^\pm \pm \sigma_i' b_i^\pm \pm \sigma_i' f^\pm [\{q_{ij \neq i}\}, \{\sigma_{j \neq i}\}] + \varepsilon_i^\pm(\sigma_i'). \tag{10}$$

The first term describes an individual's preference over one of the choices. This preference may come from an agent's characteristics, economic benefits or from external (not interpersonal) impact (e.g. public campaign). This is the counterpart of an 'external field' term in Ising model. The second one is the self-supporting term (inertia of agents). The third one denotes the influences of the other agents (interpersonal), and the fourth one is a random term.

In the mean-field approach,

$$U_i^\pm = \sigma_i' h_i^\pm \pm \sigma_i' b_i^\pm \pm \sigma_i' f^\pm \left[\left\{\frac{q}{N}\right\}, m\right] + \varepsilon_i^\pm(\sigma_i'). \tag{11}$$

As was stated above, the analysis of equilibria of the whole system is possible in the general case, as it is always possible to find functions $F^{Th-}(m)$, $F^{Th+}(m)$ as combinations of individual distribution functions $\{F_i^{Th-}(m)\}$, $\{F_i^{Th+}(m)\}$. We will examine here only some simple (and probably most realistic) cases in which to obtain global distributions from individuals' distributions is straightforward.

*a) Uniform $\{h_i^\pm\}$ and $\{\varepsilon_i^\pm \neq 0\}$ with identical distributions*

In the case of uniform coefficients and random terms with identical distribution, i.e. $h_i^\pm \to h_0^\pm$, $b_i^\pm \to b^\pm$, and $\varepsilon_i^\pm \to \varepsilon^\pm$ it is obvious, that all $\{F_i^{Th-}(m)\}$, $\{F_i^{Th+}(m)\}$ are identical and

$$F^{Th\pm}(m) = \frac{1}{N}\sum_i F_i^{Th\pm}(m) = F_i^{Th\pm}(m).$$

*b) Random $\{h_i^\pm\}$ and $\{\varepsilon_i^\pm \equiv 0\}$)*

In this case, $\{h_i^\pm\}$ is random with some distribution function $\rho^\pm(h)$; individual cumulative distribution functions $\{F_i^{Th-}(m)\}$, $\{F_i^{Th+}(m)\}$ are step functions (different, for different individuals), but their (global) combinations are described by the cumulative distribution function $\int_{-\infty}^{m} \rho^{Th\pm}(m')dm'$:

$$F^{Th\pm}(m) = \lim_{N\to\infty}\frac{1}{N}\int_{-\infty}^{m} dm' \sum_{i=1}^{N}\int dm_0 \delta(m'-m_0)\rho^\pm(m_0) = \lim_{N\to\infty}\frac{1}{N}\int_{-\infty}^{m} dm' \sum_{i=1}^{N}\rho^\pm(m') = \int_{-\infty}^{m} dm'\rho^\pm(m').$$

*c) Random $\{h_i^\pm\}$ and $\{\varepsilon_i^\pm \neq 0\}$ (with uniform distributions)*

Generally, the distribution of a sum of random variables may be a complicated function. In special cases, e.g. when distributions of $\{h_i^\pm\}$ and $\{\varepsilon_i^\pm\}$ are normal distributions (with any mean value and standard deviation) the resulting distribution is also a normal distribution with a properly calculated mean and standard deviation. In general, if the (uniform) probability densities of $h_i^\pm$ and $\varepsilon_i^\pm$ are denoted by $\rho^\pm(m)$ and $\tau^\pm(m)$ respectively, the cumulative distribution functions are given by:

$$F^{Th\pm}(m) = \int_{-\infty}^{m} dM \int_{-\infty}^{+\infty} dm' \rho^\pm(M)\tau^\pm(M - m')$$

In general, however, this case is much more complicated than the former two - here, we will focus on Cases a) and b).

It may be shown, that Cases a) and b) – although different on the individual level – within the mean-field approach, in probabilistic sense for large number of agents[2], are equivalent.
In Case a), extracting from utility functions (10) the first and last terms:

$$U_i = h_0 \sigma_i' + \tilde{U}_i + \varepsilon_i(\sigma_i')$$
$$\varepsilon_i(-1) - \varepsilon_i(+1) \sim F^\varepsilon(z)$$

where $\tilde{U}_i$ is the remaining of the deterministic part of utility function, one obtains a probability of state (+1) for a randomly chosen individual:

$$P(U(1) > U(-1)) = P(h_0 + \tilde{U}(1) + \varepsilon(1) > -h_0 + \tilde{U}(-1) + \varepsilon(-1)) =$$
$$= P(\varepsilon(-1) - \varepsilon(1) < 2h_0 + \tilde{U}(1) - \tilde{U}(-1)) = F^\varepsilon(2h_0 + \tilde{U}(1) - \tilde{U}(-1))$$

Replacing the above form with a utility function without a random term:

$$U = h\sigma_i' + \tilde{U},$$

but introducing distribution $F^h$ on parameter $h$ (case 2.):

$$h \sim F^h(z - h_0)$$

such that:

$$1 - F^h\left(-\frac{1}{2}z\right) = F^\varepsilon(z)$$

the probability of state (+1) for a randomly chosen individual reads:

$$P(U(1) > U(-1)) = P(h + \tilde{U}(1) > -h + \tilde{U}(-1)) = P(2h > \tilde{U}(-1) - \tilde{U}(1)) = P\left(h > \frac{\tilde{U}(-1) - \tilde{U}(1)}{2}\right) =$$
$$= 1 - F^h\left(-\left(h_0 - \frac{\tilde{U}(-1) - \tilde{U}(1)}{2}\right)\right) = F^\varepsilon(2h_0 + \tilde{U}(1) - \tilde{U}(-1))$$

This shows that within the mean field approach (some kind of) heterogeneity of the system is strictly equivalent to that where a random term is added to the utility function, i.e. assuming some irrationality of the agents. Levy [Levy 2005] posed a question about a model containing both heterogeneity and randomness. We show here, that in some special cases they are equivalent, i.e. some form of heterogeneity is equivalent to randomness. In the general case some more complicated calculations should be performed to obtain the joined distribution of both these terms, but it can be done in each case.

---

[2] Krauth [Krauth 2006] analyzes social interactions in small groups showing that the structure of equilibria differs from a large group case.

As these two cases are equivalent within the mean-field approach, we will use the formulation of Case b) in what follows. That is, we will omit the random term $\varepsilon$, and instead treat $h$ as a random variable with the mean value $h_0$.

Now we will proceed with the mean-field analysis of equilibria of the system (11) in cases a) and b). In this case, the cumulative distributions $F^{Th-}(m)$ and $F^{Th+}(m)$ can be expressed as:

$$F^{Th-}(m) = P(\sigma_i' = +1 | \sigma_i = -1) = P\left(h - b^- - f^-\left[\left\{\frac{q}{N}\right\}, m\right] > -h + b^- + f^-\left[\left\{\frac{q}{N}\right\}, m\right]\right) = $$
$$= 1 - F^h\left(b^- + f^-\left[\left\{\frac{q}{N}\right\}, m\right]\right) \quad (12a)$$

$$F^{Th+}(m) = 1 - P(\sigma_i' = -1 | \sigma_i = +1) = P\left(h + b^+ + f^+\left[\left\{\frac{q}{N}\right\}, m\right] > -h - b^+ - f^+\left[\left\{\frac{q}{N}\right\}, m\right]\right) = $$
$$= 1 - F^h\left(-b^+ - f^+\left[\left\{\frac{q}{N}\right\}, m\right]\right) \quad (12b)$$

where $F^h(x)$ is cumulative distribution function of the variable $h$.
Eqs (12a),(12b) together with the Eq. (9) provide conditions for which equilibria (stationary states) exist in the model (11), linking utility and impact functions with a threshold model.

5. Examples

Here we will show the link between our framework and other approaches to social modeling of interactions. In the first example, we reproduce results obtained by Levy [Levy 2005]. Then we proceed to demonstrate how, taking some special values of parameters and/or special forms of functions in (11), and stationary conditions (9), (12) one may obtain some well-known models.

*Levy's generalization*

Within the mean-field approach assuming the utility function of an individual $i$ (Eq. (1)) is independent of the individual's previous choice:

$$U_i(\sigma_i', \sigma_i, m, \{q_i\}, \{h_i\}, \varepsilon_i) \rightarrow U_i(\sigma_i', m, \{q_i\}, \{h_i\}, \varepsilon_i)$$

one obtains Levy's model [Levy 2005].
In this case $F^{Th+}(m) = F^{Th-}(m) = F^{Th}(m)$ and expression (9) reduces to:

$$m = 2F^{Th}(m) - 1 \quad (13)$$

subject to the following condition:

$$\frac{d}{dm}[U_i(+1) - U_i(-1)] > 0.$$

Distribution of effective thresholds reads:

$$F^{Th\,eff}(m) \equiv F^{Th}(m),$$

so that from (9b) we obtain:

$$m_+ = F^{Th}(m),$$

which is simply the expression presented by Levy.
Equilibria may be found by simple graphical analysis, i.e. by intersections of curves $y = m_+$ and $y = F^{Th}(m)$.

*Brock-Durlauf model*

Substituting into (5),(10) $h_i^+ = h_i^- = h_i$, $b^+ = b^- = 0$, $\varepsilon^+ = \varepsilon^-$ and $f^+ = f^- = \sum_{j \neq i} J_{ij} \sigma_j$ one obtains:

$$U_i(\sigma_i') = \sigma_i' h_i + v_i(\sigma_i') + \varepsilon_i(\sigma_i')$$

where

$$v_i(\sigma_i') = \sum_{j \neq i} J_{ij} \sigma_i' \sigma_j$$

which is (up to the constant) a well-known example of the binary choice with social interaction class of models, namely, the Brock-Durlauf model [Brock, Durlauf 2001b, Brock 2004, Durlauf 1999].

Brock and Durlauf assume that the random part of the utility function has the logistic distribution,

$$F_{\log}(z) = \frac{1}{1 + e^{-\beta z}}, \tag{14}$$

where β is called 'intensity of choice' in discrete choice literature [McFadden 1974, Manski and McFadden 1981]. As "social temperature" β has been proposed as an analogue to temperature in physical systems[3].

---

[3] See footnote 1 for discussion on social temperature concept; see also [Brock and Durlauf 2001b] for discussion on estimation of 'intensity of choice' from empirical data.

With this distribution, within the mean-field approach, the authors obtained a Curie-Weiss-like equation for equilibrium values. In this model there exist multiple equilibria (within some range of parameters) that are similar to the Ising model in statistical physics.

However, there is no need to assume the logistic distribution to obtain multiple equilibria. According to (9) and (12) within mean-field approximation one gets:

$$m = 2F^h[h+Jm]-1. \tag{15}$$

In the case of logistic distribution function equation (15) becomes:

$$m = \tanh(\beta h + \beta Jm), \tag{16}$$

but, as Levy already noted, in the cases of other distributions, the equilibria equation (15) may have, under some choice of model parameters, $2n+1$ solutions for an $n$-mode probability distribution $\rho(z) = F'(z)$ (see Fig. 1a).
For the fixed single-mode distribution the system has either one or three equilibria (two stable and one unstable) depending on the interplay of parameters (see Fig. 1b).

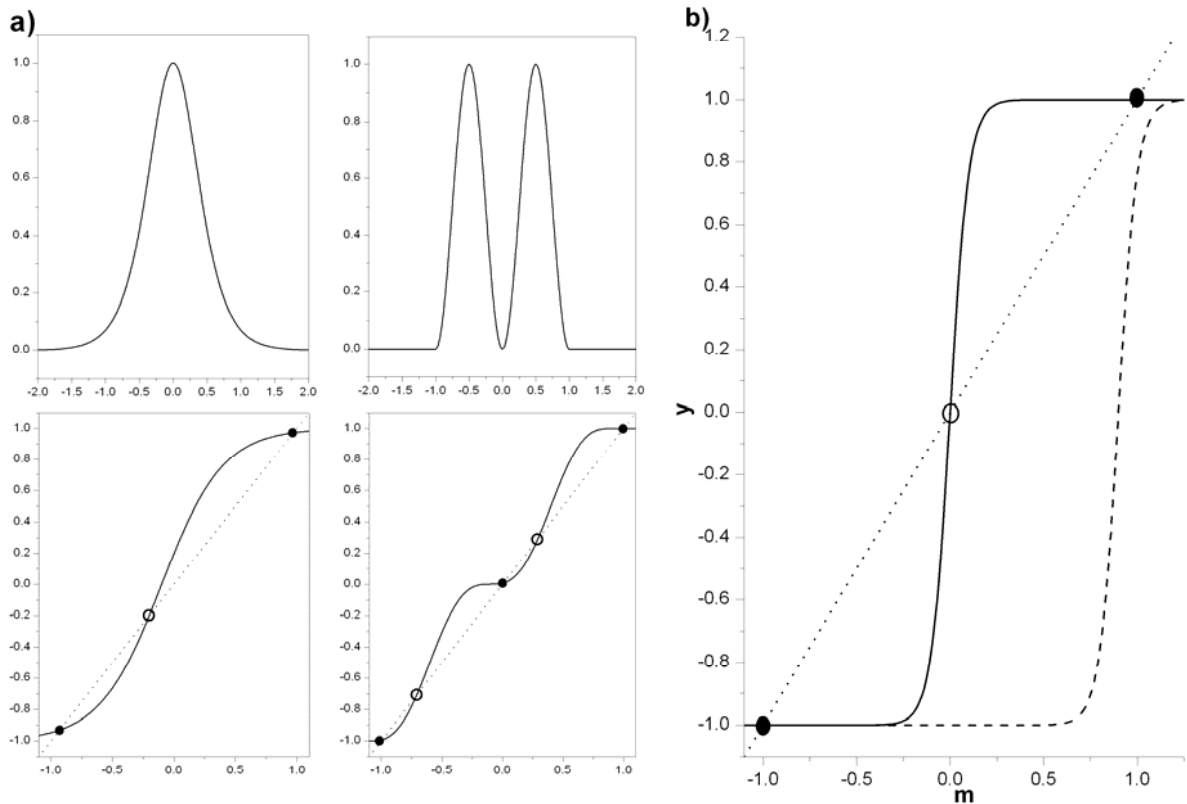

Fig.1. Graphical Analysis of the Brock-Durlauf model. a) Crossing of curves $y=m$ and $y = 2F[h+Jm]-1$ (Eq. (16)) (lower pictures) in the cases of uni- and bi-modal probability density functions (upper pictures) b) Crossing of curves $y=m$ (dotted line) and $y = 2F[h+Jm]-1$ (Eq. (15)) for $F_{\log}$ and values of parameters: $\beta = 10$, $J=2$, and $h_0 = 0$

(solid line); $h_0 = -1.8$ (dashed line). Full circles corresponds to stable stationary states and open circles to unstable stationary states.

*Holyst – Kacperski model.*

Substituting into Eqs (5) and (10) $h_i^+ = h_i^- = h_i$, $b_i^+ = b_i^- = b_i$, $\varepsilon_i^+ = \varepsilon_i^- \sim F_{\log}^\varepsilon$ and $f^+ = f^- = \sum_{j \neq i} J_{ij}\sigma_j$ one obtains:

$$U_i = \sigma_i' h_i + \sigma_i \sigma_i' b_i + \sigma_i' \sum_{j \neq i} J_{ij}\sigma_j + \varepsilon_i(\sigma_i') \tag{17}$$

which is the model of Holyst and Kacperski [Holyst *et al* 2000] (this model is a linear version of Nowak-Latane model with an additional 'external field' component).
The Holyst-Kacperski model was originally formulated using an impact function. Randomness (so-called 'social temperature') was introduced through a rule:

$$\sigma_i' = \begin{cases} \sigma_i & \text{with probability } \dfrac{1}{1+e^{2\beta I_i}} \\ -\sigma_i & \text{with probability } \dfrac{1}{1+e^{-2\beta I_i}} \end{cases},$$

where

$$I_i = -b_i - \sigma_i h - \sum_{j \neq i} J_{ij}\sigma_j.$$

It may be shown that this is strictly equivalent to the randomness introduced by adding the term $\varepsilon^\pm$ with a logistic distribution. Indeed, the probability of not-changing opinion for the utility function $U_i = -\sigma_i \sigma_i' I_i + \varepsilon(\sigma_i')$ is:

$$P(U(\sigma_i' = \sigma_i) > U(\sigma_i' = -\sigma_i)) = P(-I_i + \varepsilon(\sigma_i) > I_i + \varepsilon(-\sigma_i)) = P(\varepsilon(-\sigma_i) - \varepsilon(\sigma_i) < -2I_i) =$$
$$= \begin{cases} F^\varepsilon(-2I_i) & \text{for } \sigma_i = 1 \\ 1 - F^\varepsilon(2I_i) & \text{for } \sigma_i = -1 \end{cases}$$
.

As:

$$F^\varepsilon(-2I_i) = \frac{1}{1+e^{-\beta(-2I_i)}} = \frac{1}{1+e^{2\beta I_i}}$$
$$1 - F^\varepsilon(2I_i) = 1 - \frac{1}{1+e^{-2\beta I_i}} = \frac{1}{1+e^{2\beta I_i}},$$

this rule appears exactly the same as formulated by Holyst and Kacperski.

Within the mean-field uniform approximation ($b_i \equiv b$, $J_{ij} \equiv J$, $h_i \equiv h$) equilibria in this model can be calculated from the equation (9) with:

$$\begin{aligned} F^{Th+}(m) &= F^h(b-Jm) \\ F^{Th-}(m) &= F^h(-b-Jm) \end{aligned} \quad (18)$$

Fig. 2 shows how introducing a non-zero self-supportivness parameter may change the number of equilibria. The second equilibrium persists 'longer' (with changing parameter $h_0$) then without the self-supportivness term.

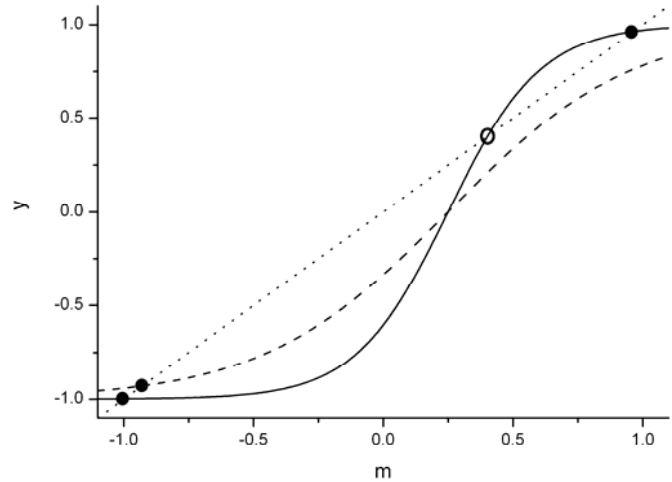

Fig.2. Graphical Analysis of the Holyst – Kacperski model. Equilibrium points are found as the intersection of curves taken from the left (dotted line) and right-hand sides of (Eqs. (9)) for $F_{\log}$ and values of parameters: $\beta = 1.5$, $J = 2$, $h_0 = -0.5$ and $b = 1$ (solid line); $b = 0$ (dashed line). Full circles corresponds to stable stationary states and open circles to unstable stationary states.

*Nowak – Latane model.*

Substituting into (5), (10) $h_i^\pm = h_i$, $b_i^+ = b_i^- = b_i$, $\varepsilon_i^+ = \varepsilon_i^-$ and

$f^\pm = -f_p\left[\sum_{j \neq i} P_{ij}(1-\sigma_i\sigma_j)\right] + f_s\left[\sum_{j \neq i} S_{ij}(1+\sigma_i\sigma_j)\right]$, where $f_p$ and $f_s$ are persuading and supporting functions, respectively, one obtains:

$$U_i = h_i\sigma_i' + \sigma_i\sigma_i'b_i - \sigma_i\sigma_i'f_p\left[\sum_j P_{ij}(1-\sigma_i\sigma_j)\right] + \sigma_i\sigma_i'f_s\left[\sum_{j \neq i} S_{ij}(1+\sigma_i\sigma_j)\right] + \varepsilon_i(\sigma_i') \quad (19)$$

which is a full version of the Nowak-Latane model [Nowak *et al* 1990, Lewenstein *et al* 1992]. Particular versions of this model omit the first and/or last term. The choice of the particular form of the functions $f_s$ and $f_p$ depends on the details of the model. In the simplest case they may be taken as linear, and this is the form considered in the authors' original papers. In this case the two terms may be easily combined into one linear tem, as in the Brock-Durlauf and Holyst-Kacperski models.

Nowak and Latane consider mostly a linear model. This model does not differ (qualitatively) from HK and BD models. Depending on the parameters' interplay, either one or three equilibria may exist.

Here we propose to examine more complex cases. Functions $f_p$ and $f_s$ do not have to be linear. More complex, non-linear forms of these functions are more realistic from the psychological perspective. For example one can notice that the first person supporting someone's opinion is much more important (adds more to agent' utility) than, for example, the eleventh person. Moreover, $f^+$ and $f^-$ do not have to be the same. That gives us four functions to specify: $f_p^+$, $f_p^-$, $f_s^+$, $f_s^-$ offering more possibilities for modeling real systems with much richer behavior obtained within various possible choices.

Below we present two examples (within mean-field uniform behavior) of the system described by (19) with $f_p^+$, $f_p^-$, $f_s^+$, $f_s^-$ modeled as Weibull functions with different parameters:

$$f_{\sup p}^+(x) = f_{\sup p}^-(x) = C_s \left[1 - \exp\left(-\left(\frac{x}{d_s}\right)^{a_s}\right)\right]$$

$$f_{pers}^+(x) = f_{pers}^-(x) = C_p \left[1 - \exp\left(-\left(\frac{x}{d_p}\right)^{a_p}\right)\right],$$

which allows a wide range of logarithmic and S-shaped functions. Random term was again described by the logistic function. Such a formulation of supporting and persuading functions offers more possibilities to adjust these functions to case study evidence. One can see that with this formulation a much richer behavior may occur, with more possible equilibria (compare Figure 3 with Figures 1 and 2) of the system.

A possible next step may be to further enrich the model by choosing different functions $f_{\sup p}^+(x) \neq f_{\sup p}^-(x)$ and $f_{pers}^+(x) \neq f_{pers}^-(x)$.

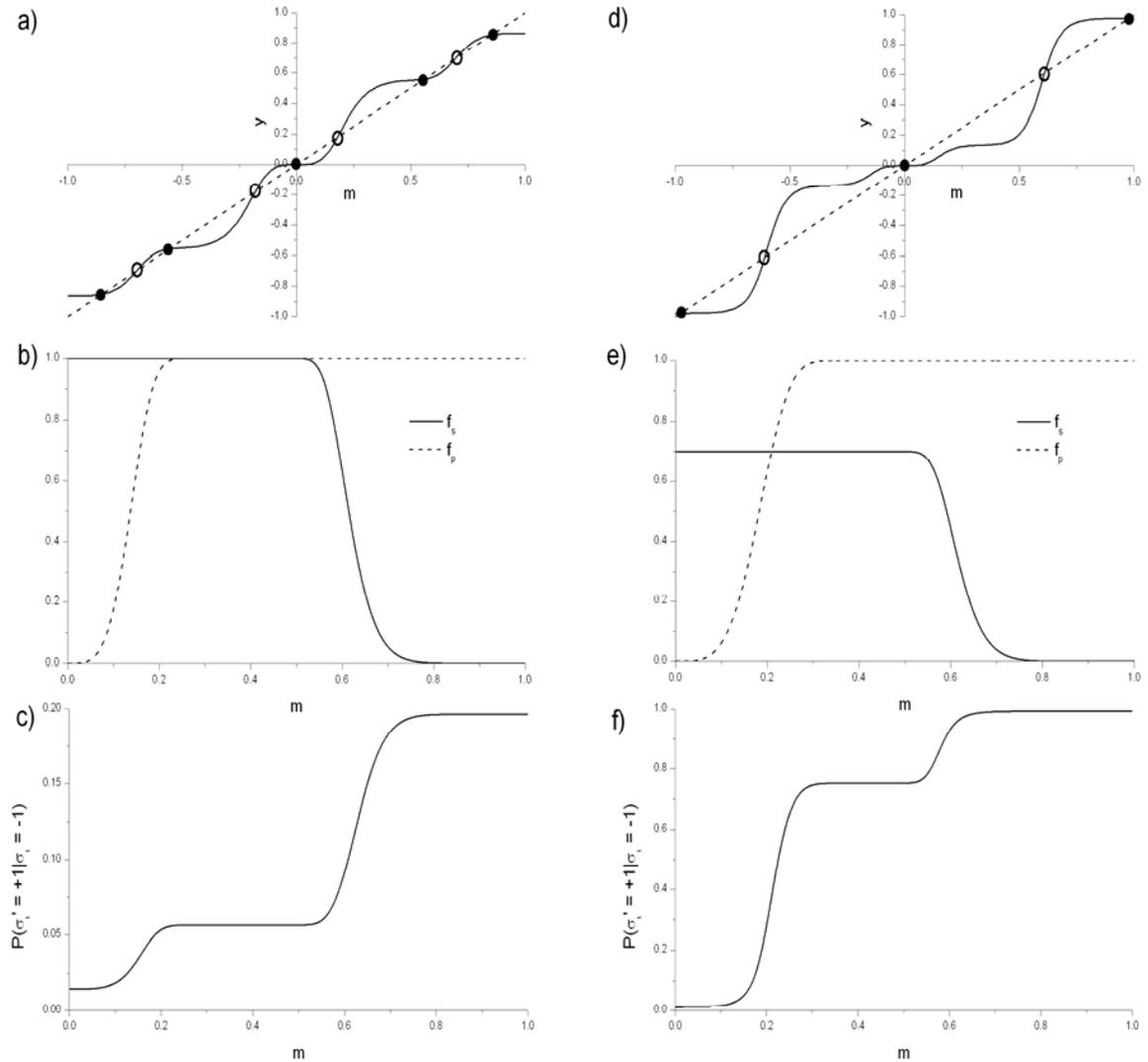

Fig. 3. Graphical Analysis of the Holyst – Kacperski model. Values of parameters: $h=0$, $b^{\pm}=2$, $P=S=1$, $C_s=C_p=1$, $d_s=0.8$, $d_p=0.3$, $a_s=10$, $a_p=4$, $\beta=1.41$ a)-c); $h=0$, $b^{\pm}=0.1$, $P=S=1$, $C_s=0.7$, $C_p=1$, $d_s=0.8$, $d_p=0.4$, $a_s=10$, $a_p=4$, $\beta=5.56$ d)-f). a) and d) show the plot of left (dashed line) and right-hand (solid line) side of Eq. (9) (full circles corresponds to stable stationary states and open circles to unstable stationary states); b) and e) show supporting and persuading functions; c) and f) show conditional probabilities of state change from (-1) to (+1)

*6. Discussion and summary*

We have discussed binary-choice models with social interactions integrating economic, sociological and psychological perspectives. The scope of analysis was restricted to the models where agents do not form expectations about others' behavior but possess direct information about others' recent choices. In this way we rule out the assumption of rational expectations, where agents possess knowledge about a system's equilibria. Also we limit agents' capabilities to form (adaptive) expectations based on the past behavior of others – the most general form of a utility function that we adopted depends only on the recent choice of other agents. This may limit application of this framework to more 'social' rather than 'market' problems. Such limitation is natural when aiming to integrate with attribute/opinion models, where agents focus on the current state of other agents rather than deliberating on their future opinions. Nevertheless, the extension of the framework to include agents' beliefs will be the subject of future work.

We have shown a close relationship between utility-function models, impact function models and threshold models. We found, that the first two of them are equivalent. In the mean field approximation, assuming certain properties of utility function, they can be reformulated as the threshold model. We have also shown, that a wide class of these models may exhibit multi-stability, and we have derived a general condition for system equilibria. It should be stressed, that within the proposed form of the utility function the expressions for equilibria hold even if the conditions for existence of thresholds are not satisfied. We have exploited a mean-field approach here. This approach may be formulated and developed in a wide variety of forms. Assuming some special form of randomness in social models, one may obtain complete analogy with statistical physics, including probability of specific states following Gibbs distribution and a principle of entropy maximizing. Such an approach was developed by Brock and Durlauf. [Brock 2004, Brock 1993, Durlauf 1997] and led to obtaining the Curie-Weiss law for stationary states of the system. With the aim to keep generality, our formulation drops out this full parallelism to physical systems, but in this special case it leads to the same results for stationary states.

We have proposed a model with a certain generalized form of utility function for more detailed analysis. This generalized model covers a wide range of models known from literature. Yet, some alternative formulations are possible. For example it is a common practice in system dynamics modeling [Sterman 1994] to model certain quantities as a product of so-called 'normal' values and certain multipliers. On the other hand recent advances in interdisciplinary research on decision-making suggest that people use very simple, heuristics for most of their decisions [Gigerenzer et al. 1999, Tversky and Kahneman 1974, McFadden 2001, Sterman 1994]. It is argued that an agent's behavior is rule-based [Axtell 2000, Moss 2002] and not utility maximizing[4]. Such perspectives can be incorporated to a certain extent here under a general individual framework (Section 2). Under this framework one can construct a model with special distributions of parameters such that different agents 'operate' according to a particular term of a utility function. For example, some agents decide based on individual preferences, other agents follow their neighbors' decisions, yet other agents decide on a habitual basis. Obviously all these elements can be present to a certain extent for different individuals. Jager et al [Jager et al. 2000] propose a model where agents whose choices depend on the uncertainty and the level of need satisfaction choose alternatively among different decision making mechanisms: deliberation, social comparison, repetition and imitation. Increasing the number of cognitive strategies available to agents is one of the directions for expansion of our framework.

It should be emphasized that we do not intend to create a general, abstract model applicable in all possible circumstances. Instead, we think that all relevant aspects of decision-making coming both from individual characteristics and the type of social interactions should be based on evidence from any particular situation. The idea of elicitation of agents' preferences, although natural in sociology, has been treated with a great dose of skepticism by economists. However, there is no sound basis to reject incorporating subjective preferences based on interviews or other forms of knowledge elicitation [Manski 2000]. The appropriate utility functions can be constructed based on agents' rules, which can be elicited using questionnaires or role-playing games [Barreteau et al. 2001].

---

[4] However, rules could evolve in such a way to satisfy underlying preferences [McFadden 2001].

One important issue of heterogeneity has been already tackled in economics – microeconometric models have departed from the 'representative agent' paradigm [McFadden 1974, Kirman 1992, Heckman 2001]. Heterogeneity of the system may be modeled in various ways, as there exist many parameters characterizing individuals. Each of them may be fixed in the whole system, or they may take different values for different agents. We show that a special kind of heterogeneity (imposed on the linear term for the 'external field') is in a simple way transformed into a model with a fixed homogenous value for the 'external field' and some randomness imposed on the choices of individuals. That is, any system with thus imposed heterogeneity may be treated as a homogenous system with some degree of randomness, and visa versa. In particular we note that this equivalence holds only within the mean-field approach with the size of the system large enough for mathematical statistics to be valid. To construct a model with both heterogeneity and randomness one has both to impose some distribution on parameters characterizing individuals and to add a random variable with some distribution to the utility function. Such a model would be still transformable to either a model with heterogeneity only or a model with randomness only, but the distribution imposed on the individuals' parameters in the first case of transformation or on random term in the second case should be calculated in a proper way, given in Section 4c. Any model with a certain distribution imposed on a random term (or 'external field term') is equivalent to all possible models with both heterogeneity and randomness, for which the combined distribution of both these terms gives this chosen distribution. That shows that taking into account both heterogeneity and randomness probably does not change qualitatively the properties of the models.

Most results in this article are obtained within the mean-field approximation. This approximation is borrowed from statistical physics. However we did not rely here on any particular statistical physics model. We constructed mean-field solutions here in such a way that they can applied to many model formulations, even those going beyond statistical mechanics[5]. It is obvious that the mean-field approach gives a strict solution in the case of complete pair-wise interactions (uniform strength). It is known from statistical physics that in the mean-field approach there always exists a phase transition. Whereas, for example, it does not happen in a one-dimensional model with finite-range interactions. The open question here is how close are the results of mean-field approximations to the results for different interaction geometries. Answering this question we can identify truly heterogeneous aspects of such models. Some effects of social networks on social interaction patterns have already been investigated (see e.g. [Valente 1996, Rolfe 2004]). Particularly interesting in this context seem to be different social network models, studied recently [Watts and Strogatz 1998, Albert and Barabasi 2002, Newman 2003]. This is a subject of our ongoing investigations.

Finally we note that models investigated here concentrate only on social dynamics and do not include a coupled environmental system. In real life actions of agents influence their environment. Environmental characteristics in turn influence agents' decisions. The framework proposed here includes environmental parameters, however it does not include any feedback between social decisions and the state of environment. Such feedback may completely change the system's equilibrium properties. It can also influence the social interactions component itself as imitation may depend on the success of observed individuals [Polhill et al. 2001]. The quest for policy relevance in environmental science has expanded the frame of inquiry to include such feedbacks, as reflected in the term *social-ecological system* (SES) [Folke 2006].

---

[5] We do not require that statistical ensemble corresponding to a specific model must have Gibbs distribution function (canonical ensemble).

**Acknowledgements**

This paper was expanded and improved as a result of questions, comments and suggestions from William Brock and Joanna Stefanska, for which we wish to extend our grateful thanks. This research was funded from EU 6$^{th}$ Framework Research Project CAVES (No 012816) and Polish Ministry of Education and Science project No 17/6.PR UE/2005/7.

25/26McFadden D., 1974. Conditional Logit Analysis of Qualitative Choice Behavior. In: Zarembka (ed), Frontiers in Econometrics. Academic Press: N.Y.

McFadden, D. 2001. Economic Choices. The American Economic Review 91(3), 351–378.

Moss, S. 2002. Policy analysis from first principles. Proceedings of the National Academy of Sciences of the United States of America 99(3), 7267–7274.

Newman M.E.J., 2003. The Structure and Function of Complex Networks. Society for Industrial and Applied Mathematics Review 45(2), 167–256.

Nowak A., Kus M., et al., 2000. Simulating the coordination of individual economic decisions. Physica A: Statistical Mechanics and its Applications 287(3-4), 613–630.

Nowak A., Lewenstein M., 1996. Modeling Social Change with Cellular Automata. Modelling and Simulation in the Social Sciences from the Philosophy of Science Point of View, 249-286.

Nowak A., Szamrej J., Latane B., From Private Attitude to Public Opinion: A Dynamic Theory of Social Impact. Psychological Review 97, 362–376.

Oliver P.E., Marwell G., 2001. Whatever Happened to Critical Mass Theory? A Retrospective and Assessment. Sociological Theory 19(3), 293-311.

Phan D., 2003. Small Worlds and Phase Transition in Agent Based Models with Binary Choices. In: Muller J.P., Seidel M.M. (Eds), 4° workshop on Agent-Based Simulation, Montpellier. SCS Publishing House, Erlangen: San Diego.

Phan D., Gordon M.B, Nadal J.P., 2004. Social Interactions in Economic Theory: an Insight from Statistical Mechanics. In: Bourgine P., Nadal J.P. (eds), Cognitive Economics, Springer Verlag, 225–358.

Platt J., 1973. Social traps. American Psychologist 28, 641–651.

Polhill J.G., Gotts N.M., Law A.N.R., 2001. Imitative Versus Non-Imitative Strategies in a Land Use Simulation. Cybernetics and Systems 32(1-2), 285–307.

Réka A., Barabási A.L., 2002. Statistical mechanics of complex networks. Reviews of Modern Physics 74, 47–97.

Rolfe M., 2004. Social networks and threshold models of collective behavior. Preprint, University of Chicago.

Rosser Jr. J.B., 1999. On the Complexities of Complex Economic Dynamics. Journal of Economic Perspectives 13(4), 169-192.

Sampson R.J., Morenoff J. D., et al., 2002. Assessing Neighborhood Effects: Social Processes and New Directions in Research. Annual Review of Sociology: 443–479.